\documentclass[12pt,draftclsnofoot,onecolumn]{IEEEtran} 
\usepackage{graphicx,cite,enumitem,}             		
			
\usepackage{amssymb,amsfonts,amsthm}   

\usepackage[cmex10]{amsmath}
\usepackage[unicode,colorlinks, linkcolor=black,anchorcolor=black,citecolor=black]{hyperref}

\def \Re {\text{Re}}
\def \Im {\text{Im}}
\def \tr {\text{tr}}

\usepackage{mathtools}

\usepackage{algorithm,caption}
\usepackage[noend]{algpseudocode}

\begin{document}

\title{Robust Receiver Design for Non-orthogonal Multiple Access}

\author{Kun~Wang \\ \today
}

\maketitle

\begin{abstract}
Non-orthogonal multiple access (NOMA) has been proposed for massive connectivity in future generations of wireless communications.
A dominant NOMA scheme is based on power optimization, in which decoding of target user is assumed to be perfect. 
In this work, rather than optimize on power domain, we are aimed to propose a robust receiver that can detect and decode 
the data streams of target user by FEC code diversity. 
Compared with existing NOMA receivers, our novel receiver substantially improves the bit-error-rate (BER) performance, 
and BER flooring can even be eliminated by assigning proper interleaving patterns to different users. 
\end{abstract}


\section{Introduction}
Most state-of-the-art cellular communication systems are utilizing orthogonal multiple access (OMA) techniques,
in which the radio resources are orthogonally allocated to multiple users. 
Notable examples of OMA techniques include the classic FDMA and TDMA, the CDMA in 3G era, and the most recent OFDMA for LTE/LTE-A. 
The orthogonality between mobile users avoids intra-cell interference and therefore simplifies system design.
Nonetheless, the number of simultaneously served users is dictated by the available orthogonal resources.

In contrast, combining the concepts of superposition coding at transmitter and successive interference cancelation (SIC) at receiver, 
non-orthogonal multiple access (NOMA) has been proposed as a promising multiple access technique for future wireless technologies 
to meet the ever-increasing demand on spectrum efficiency \cite{benjebbour2013concept}.
In NOMA, multiple users are served simultaneously on the same radio frequency. 
From an information-theoretic perspective, the NOMA scheme is optimal 
in the sense of achieving the capacity region of downlink broadcast channel \cite{tse2005fundamentals}.
Further, experimental tests of a two-user downlink NOMA cluster showed that NOMA outperforms OMA 
in terms of aggregate as well as individual user's throughput \cite{benjebbour2015non}. 

However, the increased throughput comes at the cost of severe intra-cell interferences among mobile users.
To date, most of the research investigations on NOMA assume perfect SIC at the receivers \cite{tabassum2016non}.
The practical performance of NOMA system, on the other hand, heavily relies on the successful decoding
of strong interfering signals. 
If any of these signals are decoded erroneously, the decoding errors will be propagated to subsequent stages.  
As such, it is of paramount importance to design superior receivers other than the classic SIC.
We propose to explore the underlying structure of FEC codes to strengthen receiver performance. 
Specifically, mobile users within a cell are assigned with different FEC codes or different permutations of same FEC code.
The receivers at user equipment (UE) or base station (BS) utilize the different codes or permutations
as unique user signatures to perform decoding with much improved success rates.

To fully take advantage of FEC code, we not only use FEC in decoding stage, but also make use of FEC code information in detection and demodulation.
Specifically, we will integrate the relaxed code constraints \cite{feldman2005using} to symbol detector in real/complex domain.
This set of code constraints have been widely used in our works \cite{wang2017galois}: 
space-time code as outer code is concatenated with LDPC code as inner code in \cite{wang2014joint,wang2015joint},
detection and demodulation with partial channel information is treated in \cite{wang2015diversity,wang2016diversity,wang2018unified},
turbo receiver approaching-ML performance is investigated in \cite{wang2018non,wang2018iterative,wang2018integrated},
and moreover, the asymmetry property of a class of LDPC code has been explored in \cite{wang2018semidefinite} to resolve the phase ambiguity.

\section{Decoding of FEC Codes on Real Field}
In most receiver design schemes, detection and decoding are separated as two sequential steps 
-- output of the detector is fed to the downstream decoder.
The major obstacle of combining the two steps lies in the fact that 
detection is performed on real/complex field, whereas decoding is on finite field. 
In recognition of this barrier, the quasi-joint turbo receiver was proposed to exchange 
extrinsic information iteratively between detector and decoder \cite{hochwald2003achieving}.
In the recent decade, linear-programming (LP) decoding of linear block codes 
gains wide popularity since the seminal work by Feldman \cite{feldman2005using}. 
The LP decoding opens the door for a new era of receiver design
that integrates detection and decoding stages. 

\subsection{Binary LDPC Code}
Among linear block codes, low-density parity-check (LDPC) code shows 
the capacity-approaching capability. 
An LDPC code $\mathcal{C}$ with parity check matrix $\mathbf{P}=[P_{i,j}]$ can
be represented by a Tanner graph $\mathcal{G} =
(\mathcal{V}, \mathcal{E})$.  Let $\mathcal{I} = \{1, 2, \ldots,
\emph{m}\}$ and ${\cal J} = \{1, 2, \ldots, {n}\}$,
respectively, be the row and column indices of $\mathbf{P}$. The
node set $\mathcal{V}$ can be partitioned into two disjoint node
subsets indexed by $\mathcal{I}$ and $\mathcal{J}$, known as the check nodes
and variable nodes, respectively. For each pair $(i , j) \in
\mathcal{I} \times \mathcal{J}$, there exists an edge (\emph{i},
\emph{j}) in $\mathcal{G}$ if and only if $P_{ij} = 1$. The index set of the
neighborhood of a check node $i \in \mathcal{I}$ is defined as
$\mathcal{N}_{i} := \{j \in \mathcal{J}: P_{i,j} = 1\}$. 
For each $i \in \mathcal{I}$, define the \emph{i}-th local code as
\vspace*{-1mm}
\begin{equation*}
\mathcal{C}_{i} = \{(c_{j})_{j \in \mathcal{J}}: \sum_{j
\in\mathcal{N}_{i}} P_{i,j}c_{j} = 0  \mbox{ in GF(2)}\}
\vspace*{-1mm}
\end{equation*}
where addition and multiplication are over \emph{GF}(2). 
Hence, a length-$n$ codeword
$\boldsymbol{c} \in \mathcal{C}$ if and only if 
$\boldsymbol{c} \in \mathcal{C}_i, \forall i \in \mathcal{I}$.
Therefore, decoding essentially needs to determine the most
likely binary vector $\mathbf{c}$ such that 
\vspace*{-1mm}
\begin{equation*}
\mathbf{P}\cdot \mathbf{c} = \mathbf{0}  \; \mbox{over GF(2)} 
\quad \mbox{or} \quad \Sigma_j P_{i,j}c_{j} = 0, \; \forall i\in {\cal I}.
\vspace*{-1mm}
\end{equation*}

Of interest are subsets $\mathcal{S} \subseteq \mathcal{N}_i$ that contain an even number of variable nodes; 
each such subset corresponds to a local codeword \cite{yang2006nonlinear}.
Let $\mathcal{E}_i \triangleq \{ \mathcal{S} \, | \, \mathcal{S} \subseteq \mathcal{N}_i \, \text{with} \, |\mathcal{S}| \, \text{even} \}$, and introduce auxiliary variable $v_{i,\mathcal{S}} \in \{0, 1\}$ to indicate the local codeword associated with $\mathcal{S}$. 
Since each parity check node can only be satisfied with one particular even-sized subset $\mathcal{S}$,  
the following equation must hold \cite{feldman2005using}
\vspace*{-1mm}
\begin{equation} \label{eq:code_v}
\sum_{\mathcal{S} \in \mathcal{E}_i} v_{i,\mathcal{S}} = 1, \; \forall i \in \mathcal{I}.
\vspace*{-1mm}
\end{equation}
Moreover, use $f_j \in \{0, 1\}$ to represent variable node $j$, indicating a bit value of 0 or 1.
The bit variables $f_j$'s must be consistent with each local codeword. Thus,
\vspace*{-1mm}
\begin{equation} \label{eq:code_f}
\sum_{\mathcal{S} \in \mathcal{E}_i: j \in \mathcal{S}} v_{i,\mathcal{S}} = f_j, \; \forall j \in \mathcal{N}_i, i \in \mathcal{I}.
\vspace*{-1mm}
\end{equation}

To see how these code constraints characterize a valid codeword at the $i$-th parity check, 
note that, according to constraint (\ref{eq:code_v}) and the fact that $v_{i,\mathcal{S}}$ takes integer values, 
we have $v_{i,\mathcal{S}'} = 1$ for some $\mathcal{S}'$ and $v_{i,\mathcal{S}''} = 0$ for all other $\mathcal{S}'' \neq \mathcal{S}'$,
where $\mathcal{S}', \mathcal{S}'' \in \mathcal{E}_i$.
Furthermore, from constraint (\ref{eq:code_f}), we have $f_j = 1$ for all $j \in \mathcal{S}'$ and 
$f_j = 0$ for all $j \in \mathcal{N}_i \backslash \mathcal{S}'$. 
Since $|\mathcal{S}'|$ is even-sized, the $i$-th parity check is satisfied.
Constraints (\ref{eq:code_v}) and (\ref{eq:code_f}) are enforced for every parity check. 
Together they define a valid codeword \cite{feldman2005using}. 
Notice that the constraint $v_{i,\mathcal{S}} \in \{0, 1\}$ would lead to integer programming, which is computationally expensive. 
Therefore, it is relaxed to $0 \leq v_{i,\mathcal{S}} \leq 1$. 
Meanwhile, constraint (\ref{eq:code_f}) guarantees that $0 \leq f_j \leq 1$.

The decoding constraints (\ref{eq:code_v}) and (\ref{eq:code_f}) use exponentially many variables $\{ v_{i,\mathcal{S}} \}$.
On the other hand, the constraints can be exponentially many, while with only $n$ variables $\{ f_i \}$.
This time, let $\mathcal{S} \triangleq \{ \mathcal{F} \, | \, \mathcal{F} \subseteq \mathcal{N}_i \, \text{with} \, |\mathcal{F}| \, \text{odd} \}$.
The fundamental polytope characterizing code property is captured by 
the following forbidden set (FS) constraints \cite{feldman2005using}
\begin{equation} \label{eq:parity_ineq}
\sum_{ i \in \mathcal{F} } f_i - \sum_{ i \in \mathcal{N}_i \backslash \mathcal{F}} f_i \leq |\mathcal{F}| - 1, \; \forall i \in \mathcal{I},
\forall \mathcal{F} \in \mathcal{S}
\end{equation}
plus the box constraints for bit variables
\begin{equation} \label{eq:box_ineq}
 0  \leq f_i \leq 1, \quad \forall i \in \mathcal{I}.
\end{equation}
In fact, if the variables $f_i$'s are zeros and ones, these constraints will be equivalent 
to the original binary parity-check constraints.
To see this, if parity check node $i$ fails to hold, there must be a subset of variable nodes
$\mathcal{F} \subseteq \mathcal{N}_i$ of odd cardinality such that all nodes in $\mathcal{F}$
have the value 1 and all those in $\mathcal{N}_i \backslash \mathcal{F}$ have value 0.
Clearly, the corresponding parity inequality in (\ref{eq:parity_ineq}) would forbid this situation.

\subsection{High-density code: Polar and Reed Muller}
The number of variables $v_{i,\mathcal{S}}$ in Eq.~(\ref{eq:code_v}) is exponential 
in the degree of check node. 
Similarly, the number of constraints in Eq.~(\ref{eq:parity_ineq}) is also exponential.
Thus, for codes with high-density parity-check matrices, these formulations
are computationally expensive.
Alternatively, the works \cite{yang2008new} and \cite{chertkov2007pseudo} showed an LP formulation
 whose size is linear in the code length and check node degree.
 The formulation is obtained through a decomposition approach:
 a high-degree check node is decomposed into several low-degree check nodes
 by adding auxiliary variable nodes. 
 This technique can be applied iteratively to check nodes 
 until each check node is associated with two or three variable nodes.
 For a check node with three variable nodes $f_1$, $f_2$ and $f_3$,
 the parity check constraint $f_1 + f_2 + f_3 = 0$ (mod 2) can be 
relaxed to a set of linear constraints 
\vspace{-3mm}
\begin{subequations} \label{eq:decomp_poly}
\begin{align}
0 \leq f_1 \leq f_2 + f_3, \\
0 \leq f_2 \leq f_3 + f_1, \\
0 \leq f_3 \leq f_1 + f_2, \\
f_1 + f_2 + f_3 \leq 2, \\
0 \leq f_1, f_2, f_3 \leq 1, 
\vspace{-3mm}
\end{align}
\end{subequations}
For a parity check node with two variable node, the constraint is simply
$0 \leq f_1 = f_2 \leq 1$.

In recent years, polar code was discovered to be capacity-achieving
on symmetric binary-input discrete memoryless channels 
such as the binary symmetric channel (BSC) and binary erasure channel (BEC) \cite{arikan2009channel}.
Let $\mathbf{G}_2^{\otimes n} = \mathbf{G}_2 \otimes \cdots \otimes \mathbf{G}_2$ be the $n$-fold
Kronecker product of the polarizing kernel $\mathbf{G}_2 = \left[ \begin{smallmatrix}  1 & 0 \\ 1 & 1  \end{smallmatrix} \right]$. 
Then, the polar codes are encoded as $\mathbf{c} = \mathbf{u} \mathbf{G}_N$ via the generator matrix
$\mathbf{G}_N = \mathbf{B}_N \mathbf{G}_2^{\otimes n}$, in which $\mathbf{B}_N$ is a bit-reversal permutation matrix.
The factor graph representation of  $\mathbf{G}_N$ consists of 2-degree and 3-degree check nodes only.
Thus, the above parity check constraints in Eq.~(\ref{eq:decomp_poly}) can naturally be applied \cite{goela2010lp}.
The capacity-approaching performance of polar code is proven under the assumption of infinitely long code.
For practically long code, a similar code, Reed Muller (RM) code, is ``re-discovered'' with superior performance. 
The full RM generator matrix takes the form $G_{RM}(n,n) = \mathbf{G}_2^{\otimes n}$, 
while the $r$-th order RM code $RM(r, n)$ can then be defined as the linear code with generator matrix $G_{RM}(r, n)$
 which is obtained by taking the rows of $G_{RM}(n, n)$ with Hamming weights $\geq 2^{n-r}$ \cite{arikan2008performance}.
 Recognizing the similarity between polar and RM codes, we can also apply the decomposed 
 parity check constraints to RM codes of short to medium length.

\subsection{Beyond Linear Block Codes}
So far, our focus is on linear block codes, including low density and high density codes. 
The common characteristic of these LP formulations is that they are derived based on 
the property of parity checks. 
Another large class of codes, such as convolutional codes and turbo codes, 
are encoded based on finite state machine. They can naturally be represented by a trellis. 
Accordingly, linear programming decoder for turbo-like codes was proposed in \cite{feldman2004decoding}.
In this LP formulation, the trellis graph of each constituent encoder $C^{\nu}$
is modeled by flow conservation and capacity constraints \cite{ahuja2017network}.
Moreover, extra constraints are needed to connect the flow variables with codeword variables. 

First comes the flow constraints. Let $G_{\nu} = (S_{\nu}, E_{\nu})$ be the trellis according to $C_{\nu}$,
where $S_{\nu}$ is the index set of states and $E_{\nu}$ is the set of edges (denoting state transitions)
in $G_{\nu}$. Moreover, $s_{\nu}^{start}$ and $s_{\nu}^{end}$ represent starting state and ending state,
respectively. Then, a feasible flow is characterized
\vspace{-2mm}
\begin{subequations}
\begin{align}
\sum_{e \in out(s_{\nu}^{start})} f_e^{\nu} = 1, \; \sum_{e \in in(s_{\nu}^{end})} f_e^{\nu} = 1, \\
\sum_{e \in out(s)} f_e^{\nu} = \sum_{e \in in(s)} f_e^{\nu}, \; \forall s \in S_{\nu} \backslash \{s_{\nu}^{start}, s_{\nu}^{end}  \}, \\
f_e^{\nu} \geq 0, \; \forall e \in E_{\nu}.
\end{align}
\end{subequations}
Further, the connecting constraints between flow variables and code variables are 
\vspace{-2mm}
\begin{equation}
x_j^{\nu} = \sum_{e \in O_j^{\nu}} f_e^{\nu}, \; \nu \in \{a,b\}; \quad 
x_j^s = \sum_{e \in I_j^a} f_e^a; \quad x_{\pi(j)}^s = \sum_{e \in I_j^b} f_e^b,
\end{equation}
where $x_j^{\nu}$ and $x_j^s$ embody the turbo codeword, and $\pi(\cdot)$ denotes interleaving.
This formulation can be generalized to all kinds of ``turbo-like'' codes that are built on 
convolutional codes plus interleavers.

\section{Code-diverse Receiver in NOMA}

\begin{figure}[!ht]
\begin{center}
\includegraphics[scale=0.5]{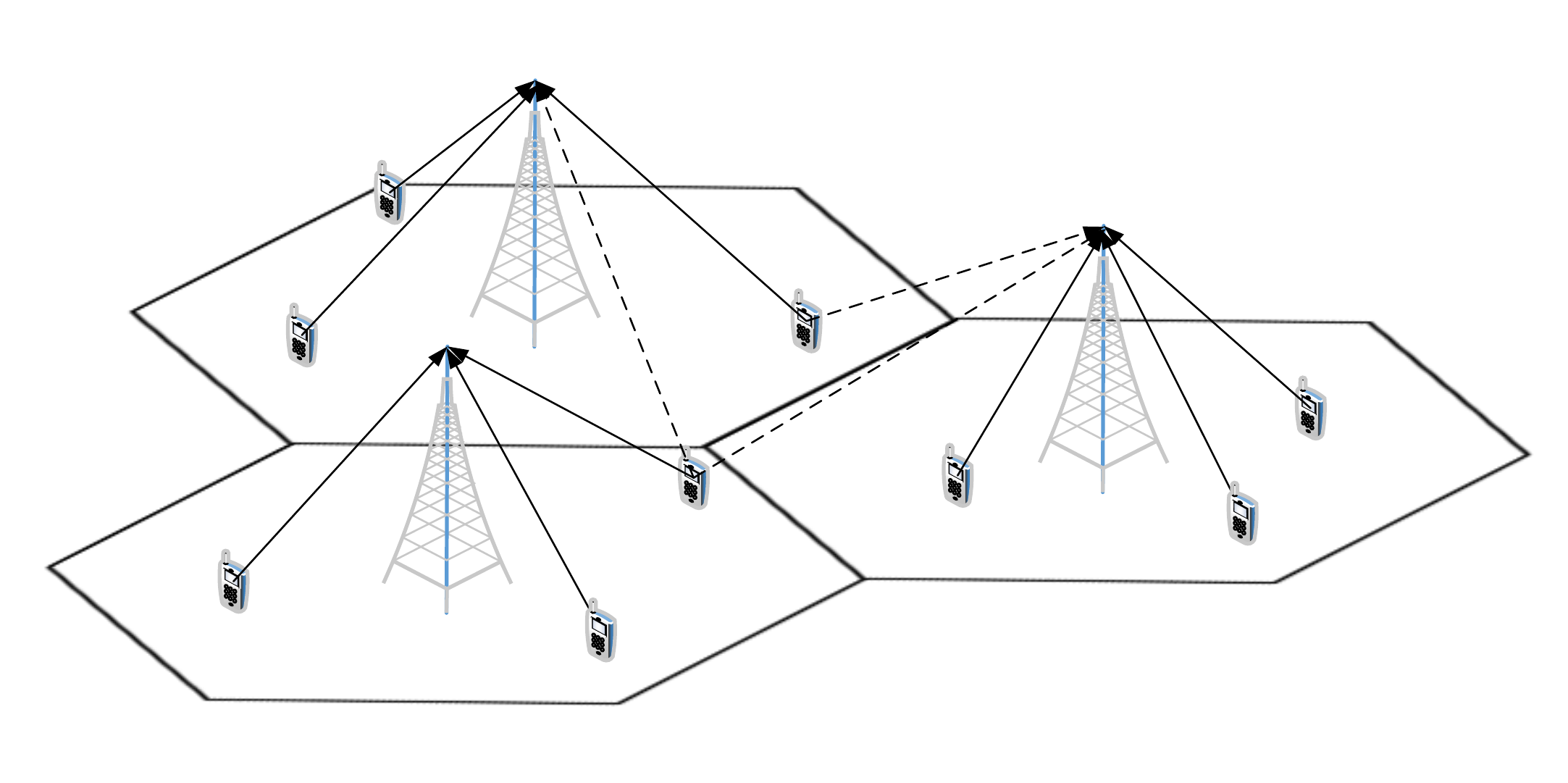}
\vspace*{-3mm}
\caption{ Uplink transmissions in an MC-MU-MIMO system. } \label{fig:mimo_sys}
\vspace*{-3mm}
\end{center}
\end{figure}

We consider the newly proposed NOMA communication scenario,
consisting of a central base station (BS) and multiple user equipments (UEs).
The UEs send signals in overlapped time-frequency resources,
such that the signals at receiver interfere with each other.
In modern cellular deployment, because of aggressive frequency reuse, 
inter-cell signals cause interferences to received signals
besides intra-cell interferences. 
In Fig.~\ref{fig:mimo_sys}, we show an example of uplink transmissions 
in a multi-cell (MC) multi-user (MU) MIMO system. 
The dashed lines represent such inter-cell interferences.
Similarly, in the downlink direction, UEs receive interferences
from BS's broadcasting signals, even though transmit beamforming is performed at BS. 

To mitigate interference, classic successive interference cancelation (SIC) receiver is widely used. 
One critical step for the success of SIC is the cancelation order. 
However, it is difficult to precisely determine the order.  
Instead, we propose FEC code diversity to facilitate the SIC process. 
The form of diversity can be different FEC codes or same FEC codes but different permutations. 
By assigning each UE with a unique FEC code or permutation, the UE's signal 
can be extracted with much improved accuracy. 
In the following, we demonstrate some preliminary works on receive beamforming in uplink direction.

The uplink transmission in a multi-cell multi-user MIMO (MC-MU-MIMO) network 
that consists of $L$ cells is considered.
The $\ell$-th cell BS serves $K_{\ell}$ single-antenna mobile users,
and every BS is equipped with $N_r$ antennas.
The flat-fading uplink channel matrix from all $K_{\ell}$ users in the $\ell$-th cell to the $i$-th BS
is represented by $\tilde{\mathbf{H}}_{i,\ell}$.
The complex-valued channel coefficients in $\tilde{\mathbf{H}}_{i,\ell}$ comprise the effects of
both small-scale fading and large-scale attenuation.
In order to facilitate the subsequent derivations, symbols in complex field are transformed to real field. 
Received signal $\mathbf{y} = [ \Re\{\tilde{\mathbf{y}}\}^T \; \Im\{\tilde{\mathbf{y}}\}^T ]^T \in \mathbb{R}^{2N_r \times 1}$;
transmitted signal $\mathbf{x} = [\mathbf{x}_1^T \ldots \mathbf{x}_{N_u}^T]^T \in \mathbb{R}^{2N_u \times 1}$ 
where the $k$-th user's transmitted signal $\mathbf{x}_{k} = [\Re\{\tilde{x}_k\} \; \Im\{\tilde{x}_k\}]^T \in \mathbb{R}^{2 \times 1}$;
Gaussian random noise $\mathbf{n} = [ \Re\{\tilde{\mathbf{n}}\}^T \; \Im\{\tilde{\mathbf{n}}\}^T ]^T \in \mathbb{R}^{2N_r \times 1}$;
channel matrix $\mathbf{H} = [\mathbf{H}_1 \ldots \mathbf{H}_{N_u}] \in \mathbb{R}^{2N_r \times 2N_u}$, in which the $k$-th user's channel matrix is $\mathbf{H}_{k}$.
Finally, the real-valued signal transmit-receive equation is $\mathbf{y} = \mathbf{H} \mathbf{x} + \mathbf{n}$.

Signal detection at receiver heavily relies on channel estimate. 
However, practical cellular systems are implementing 
more and more aggressive frequency reuse and denser cell deployment.
One consequence is that the target user's channel estimate based on pilot training is perturbed
by channel coefficients of other interfering users. 
This phenomenon is called \textit{pilot contamination} \cite{marzetta2010noncooperative}. 
Especially in the massive MIMO systems that target to serve many more mobile users, 
the effect of pilot contamination can exacerbate \cite{larsson2014massive,rusek2013scaling}.
Mathematically, when users in nearby cells use the same set of training sequences synchronously, 
the resulting estimated channel matrix of user $k$ becomes
$ \hat{\mathbf{H}}_k = \mathbf{H}_k + \sum_{j \in \mathcal{I}} \mathbf{H}_{j} + \hat{\mathbf{N}},$
where $\mathcal{I}$ is the set of interfering users during training phase and
 $\hat{\mathbf{N}}$ is the error term because of noise
associated with channel estimates \cite{lu2014overview,krishnan2014uplink}.
Users in set $\mathcal{I}$ are called the \textit{pilot-interfering users}.

Without loss of generality, we designate the first user as the target user.
Based on the MOE criterion, a minimum variance (MV) receiver in the context of multi-access MIMO systems was proposed in \cite{shahbazpanahi2004minimum}. Specifically, the MV detector is the solution to the following optimization problem
\begin{equation} \label{eq:dlmv}
\begin{aligned}
& \underset{\mathbf{W}}{\text{min.}}
& & \tr \{ \mathbf{W}^T \mathbf{R} \mathbf{W} \}  \\
& \text{s.t.}
& & \mathbf{W}^T \hat{\mathbf{H}}_1 = \mathbf{I}_2,
\end{aligned}
\end{equation}
where $\mathbf{R} = \mathbb{E}\{ \mathbf{y} \mathbf{y}^T \} \in \mathbb{R}^{2N_r \times  2N_r}$ is the covariance matrix. 
In practice, the unknown covariance matrix $\mathbf{R}$ is replaced by 
its estimate from $T_R$ snapshots of received signals
$\hat{\mathbf{R}} = \frac{1}{T_R} \sum_{t=1}^{T_R} \mathbf{y}_t \mathbf{y}_t^T$.
Furthermore, to provide additional robustness against finite samples and to strengthen the 
condition number of $\hat{\mathbf{R}}$, diagonally loaded covariance matrix
$\check{\mathbf{R}} = \hat{\mathbf{R}} + \gamma \mathbf{I}$ is used to replace $\hat{\mathbf{R}}$,
where $\gamma$ is the diagonal loading (DL) factor.

Let $\mathbf{w} \triangleq [\mathbf{w}_R^T \; \mathbf{w}_I^T]^T$ be the vectorized  
receiver parameter matrix $\mathbf{W}$. We then rewrite the MOE cost function in quadratic form
\vspace*{-2mm}
\begin{equation} \label{eq:qp_cost}
\tr\{ \mathbf{W}^T \check{\mathbf{R}} \mathbf{W} \}  =
\begin{bmatrix}
\mathbf{w}_R^T & \mathbf{w}_I^T
\end{bmatrix}
\begin{bmatrix}
\check{\mathbf{R}} &  \\
                  & \check{\mathbf{R}} 
\end{bmatrix}
\begin{bmatrix}
\mathbf{w}_R \\
 \mathbf{w}_I
\end{bmatrix}.
\vspace*{-2mm}
\end{equation}
For the constraint, we follow the same vectorization strategy
\vspace*{-2mm}
\begin{equation} \label{eq:zf_constr}
\mathbf{W}^T \hat{\mathbf{H}}_1 = \mathbf{I}_2 \; \Longleftrightarrow 
\begin{bmatrix}
\hat{\mathbf{H}}_1^T &  \\
                  & \hat{\mathbf{H}}_1^T 
\end{bmatrix}
\begin{bmatrix}
\mathbf{w}_R \\
 \mathbf{w}_I
\end{bmatrix}
=
\begin{bmatrix}
\mathbf{e}_1 \\
\mathbf{e}_2
\end{bmatrix},
\end{equation}
where the unit vectors $\mathbf{e}_1 = [1 \;\; 0]^T$ and $\mathbf{e}_2 = [0 \;\; 1]^T$ constitute the identity matrix $\mathbf{I}_2$.

Since CSI mismatch exists, we would like to enforce less stringent constraint on response preservation. 
Particularly, we lift the interference residual $\Vert (\mathbf{I}_2 \otimes  \hat{\mathbf{H}}_1^T) \mathbf{w}  - \mathbf{e} \Vert^2$, where $\mathbf{e} \triangleq [\mathbf{e}_1^T \;\; \mathbf{e}_2^T]^T$,
into the cost function to arrive at an unconstrained QP formulation with regularization parameter $\alpha$
\vspace*{-2mm}
\begin{equation} \label{eq:unconstr_qp}
\underset{\mathbf{w}}{\text{min.}} \;
\mathbf{w}^T (\mathbf{I}_2 \otimes \check{\mathbf{R}}) \mathbf{w} + \alpha \Vert (\mathbf{I}_2 \otimes  \hat{\mathbf{H}}_1^T) \mathbf{w}  - \mathbf{e} \Vert^2.
\vspace*{-2mm}
\end{equation}

To accomplish the integration of code constraints (\ref{eq:code_v}) and (\ref{eq:code_f}) 
into the QP receiver in Eq.~(\ref{eq:unconstr_qp}), 
we will employ additional (linear) constraints that connect the recovered symbols 
$\hat{\mathbf{x}}_1 = \mathbf{W}^T \mathbf{y}$ to the bit variables $\{f_j\}_{j \in \mathcal{J}}$.
The 4-QAM with Gray mapping admits affine relationship between bits and symbols
\vspace*{-2mm}
\begin{subequations} \label{eq:4qam_gray}
\begin{align}
\mathbf{w}_R^T \mathbf{y}_t & = (2 f_{2t-1} - 1) / \sqrt{2}, \\
\mathbf{w}_I^T \mathbf{y}_t & = (1 - 2 f_{2t}) / \sqrt{2}.
\vspace*{-5mm}
\end{align}
\end{subequations}

In order to use the code information for the purpose of distinguishing target user from pilot-interfering users, 
we can allocate different codes to different users as unique signatures \cite{wang2016robust,wang2016fec}. 
However, in practice, the number of mobile users may far exceed the channel codes specified in the system standard.
Since interleaver and deinterleaver are commonly utilized for time-diversity, 
we propose to apply different codeword permutations to different users as unique signatures.
Specifically, for the 1st user,  codeword $\mathbf{f} = [f_1, \ldots, f_{n}]$ is permuted 
by a permutation matrix $\boldsymbol{\Pi}_1$ before modulation. 
Therefore, the permuted codeword $\mathbf{f}^p = \boldsymbol{\Pi}_1 \mathbf{f}$.
Before finalizing the joint receiver, 
we point out that the bit variables $f_n$'s in bit-to-symbol mapping constraints [Eq.~(\ref{eq:4qam_gray})]
should be replaced by the permuted variables $f^p_n$'s while those in code constraints [Eqs.~(\ref{eq:code_v}) and (\ref{eq:code_f})] remain unchanged.
Finally, the joint QP receiver anchored with FEC code is summarized below with 4-QAM as a typical case 
\begin{equation} \label{eq:joint_qp}
\begin{aligned}
& \underset{\mathbf{w}, \mathbf{f}, \mathbf{v}}{\text{min.}}
& & \mathbf{w}^T (\mathbf{I}_2 \otimes \check{\mathbf{R}}) \mathbf{w} + \alpha \Vert (\mathbf{I}_2 \otimes  \hat{\mathbf{H}}_1^T) \mathbf{w}  - \mathbf{e} \Vert^2   \\
& \text{s.t.}
& & \mathbf{w}_R^T \mathbf{y}_t = (2 f^p_{2t-1} - 1) / \sqrt{2}, \\
&
& & \mathbf{w}_I^T \mathbf{y}_t = (1 - 2 f^p_{2t}) / \sqrt{2}, \\
& 
& & \mathbf{f}^p = \boldsymbol{\Pi}_1 \mathbf{f}, \\
&
& & \sum_{\mathcal{S} \in \mathcal{E}_i} v_{i,\mathcal{S}} = 1, \; \forall i \in \mathcal{I}, \\
&
& & \sum_{\mathcal{S} \in \mathcal{E}_i: j \in \mathcal{S}} v_{i,\mathcal{S}} = f_j, \; \forall j \in \mathcal{N}_i, i \in \mathcal{I}, \\
&
& & 0 \leq v_{i,\mathcal{S}} \leq 1, \forall i \in \mathcal{I}, \mathcal{S} \in \mathcal{E}_i.
\end{aligned}
\end{equation}

We note that the permutation constraint $\mathbf{f}^p = \boldsymbol{\Pi}_1 \mathbf{f}$ merely reorders vector $\mathbf{f}$ to $\mathbf{f}^p$.
For practical implementation, we do not wish to incorporate $\mathbf{f}^p = \boldsymbol{\Pi}_1 \mathbf{f}$ 
by introducing many more constraints as well as more variables (the vector $\mathbf{f}^p$), 
leading to higher computational complexity.
Instead, we just need to ensure the coefficients of the variables are in their correct storage units.

\begin{figure}[!tb]
\begin{center}
\includegraphics[width=9cm]{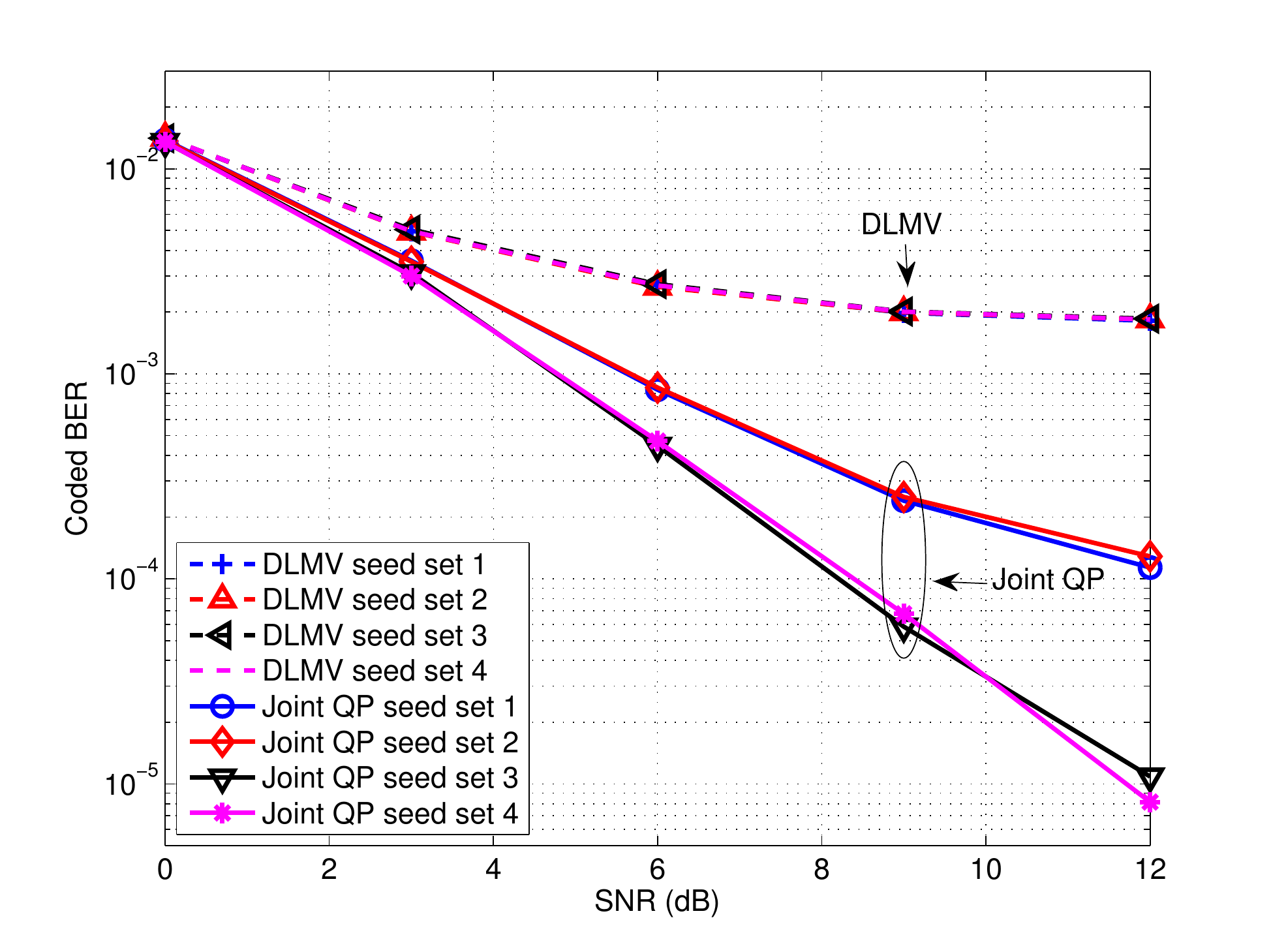}
\caption{ \small{Demonstration of code diversity effect: DLMV receiver versus joint QP receiver. 
4-QAM. $N_r=32$. LDPC code (256,192). DL factor $\gamma = 200 \sigma_n^2$ and regularization parameter $\alpha=75$. 
Three users with channel power gains = (1, 0.3, 0.7). Seed set 1 = (568,568,568), 
set 2 = (568,568,625), set 3 = (568,193,568) and set 4 = (568,193,625).} }
\vspace*{-5mm} 
\label{fig:code_div}
\end{center}
\end{figure}

We demonstrate the advantage given by FEC code diversity in Fig.~\ref{fig:code_div}.
Consider 3 interfering uplink users in the network, and their channel power gains to the BS are 
1, 0.3 and 0.7, respectively. The BS is equipped with 32 antennas.
To clearly illustrate the effect of code anchoring, we compare 
different combinations of code permutations by showing the coded BER versus signal-to-noise ratio (SNR).
The permutations are controlled by seeds, which are listed in the figure caption.

The pilot contamination exists between the first and the second user.
Hence, FEC code diversity between these two users is important to our code-anchored
QP receiver. Under permutation set 3 and set 4, the target user and pilot-interfering user possess
unique FEC permutations as signatures.   
By exploiting the FEC code diversity through code anchoring, the
BER of QP receiver under permutation set 3 and set 4 in Fig.~\ref{fig:code_div}
is reduced by as much as 3 orders of 
magnitude under pilot contamination when compared with DLMV.  The similar BER results
achieved under these two permutation sets also show that even when user 3 uses the same
FEC code and permutation,  the BER results remain unchanged because of the
lack of pilot contamination from user 3.  

On the other hand,  when the target user and pilot-interfering user share the same code permutation (in sets 1 and 2),  
the coded BER of joint QP is severely degraded, regardless of the third user's code permutation. Nevertheless,
exploiting FEC code information by the QP receiver can still improve the receiver BER even when
the users in pilot contamination are assigned the same FEC code and permutation scheme.

\section{Summary and Future Works}
Besides the above robust receiver design, there are many other optimizations to do in NOMA.
In the downlink NOMA paradigm proposed by DOCOMO \cite{benjebbour2013concept}, 
UEs are grouped in pair of two that share the same frequency band.
The two UEs should be far away from each other in distance. 
BS transmits superposed signals to the UEs in that pair, 
where signal of the farther UE uses stronger transmission power and that of the nearer UE 
has weaker power.
The consequence is the nearer UE receives very high interference from the signal of the farther UE.
Thus, this nearer UE has to first decode the farther UE's signal and cancel it out before decoding its own signal.
On the contrary, the farther UE can decode its signal directly since the interference is weak. 
One critical step in the success of this transmission paradigm is power allocation.
We should formulate a joint power and detection optimization problem that can 
achieve best decoding performance at both UEs with optimized transmission powers.
This kind of optimization might be performed alternately between power optimization and detection optimization.
Besides joint power and detection optimization, we can also try to apply the technique to jointly design with precoder \cite{wu2014cooperative,wu2015cooperative}.
In addition, the joint receiver design is useful to combat with RF imperfections \cite{wang2017phase}.

\ifCLASSOPTIONcaptionsoff
  \newpage
\fi



%

\bibliographystyle{IEEEtran}
\bibliography{IEEEabrv,mybibfile}

\begin{thebibliography}{10}
\providecommand{\url}[1]{#1}
\csname url@samestyle\endcsname
\providecommand{\newblock}{\relax}
\providecommand{\bibinfo}[2]{#2}
\providecommand{\BIBentrySTDinterwordspacing}{\spaceskip=0pt\relax}
\providecommand{\BIBentryALTinterwordstretchfactor}{4}
\providecommand{\BIBentryALTinterwordspacing}{\spaceskip=\fontdimen2\font plus
\BIBentryALTinterwordstretchfactor\fontdimen3\font minus
  \fontdimen4\font\relax}
\providecommand{\BIBforeignlanguage}[2]{{%
\expandafter\ifx\csname l@#1\endcsname\relax
\typeout{** WARNING: IEEEtran.bst: No hyphenation pattern has been}%
\typeout{** loaded for the language `#1'. Using the pattern for}%
\typeout{** the default language instead.}%
\else
\language=\csname l@#1\endcsname
\fi
#2}}
\providecommand{\BIBdecl}{\relax}
\BIBdecl

\bibitem{benjebbour2013concept}
A.~Benjebbour, Y.~Saito, Y.~Kishiyama, A.~Li, A.~Harada, and T.~Nakamura,
  ``Concept and practical considerations of non-orthogonal multiple access
  {(NOMA)} for future radio access,'' in \emph{Intelligent Signal Processing
  and Communications Systems (ISPACS), 2013 International Symposium on}.\hskip
  1em plus 0.5em minus 0.4em\relax IEEE, 2013, pp. 770--774.

\bibitem{tse2005fundamentals}
D.~Tse and P.~Viswanath, \emph{Fundamentals of wireless communication}.\hskip
  1em plus 0.5em minus 0.4em\relax Cambridge university press, 2005.

\bibitem{benjebbour2015non}
A.~Benjebbour, K.~Saito, A.~Li, Y.~Kishiyama, and T.~Nakamura, ``Non-orthogonal
  multiple access {(NOMA)}: Concept, performance evaluation and experimental
  trials,'' in \emph{Wireless Networks and Mobile Communications (WINCOM), 2015
  International Conference on}.\hskip 1em plus 0.5em minus 0.4em\relax IEEE,
  2015, pp. 1--6.

\bibitem{tabassum2016non}
H.~Tabassum, M.~S. Ali, E.~Hossain, M.~Hossain, D.~I. Kim \emph{et~al.},
  ``Non-orthogonal multiple access (noma) in cellular uplink and downlink:
  Challenges and enabling techniques,'' \emph{arXiv preprint arXiv:1608.05783},
  2016.

\bibitem{feldman2005using}
J.~Feldman, M.~J. Wainwright, and D.~R. Karger, ``Using linear programming to
  decode binary linear codes,'' \emph{Information Theory, IEEE Transactions
  on}, vol.~51, no.~3, pp. 954--972, 2005.

\bibitem{wang2017galois}
K.~Wang, ``Galois meets {E}uclid: {FEC} code anchored robust design of wireless
  communication receivers,'' Ph.D. dissertation, University of California,
  Davis, 2017.

\bibitem{wang2014joint}
K.~Wang, W.~Wu, and Z.~Ding, ``Joint detection and decoding of {LDPC} coded
  distributed space-time signaling in wireless relay networks via linear
  programming,'' in \emph{Proc. IEEE Int. Conf. Acoust., Speech, Signal
  Process. (ICASSP), Florence, Italy}, July 2014, pp. 1925--1929.

\bibitem{wang2015joint}
K.~Wang, H.~Shen, W.~Wu, and Z.~Ding, ``Joint detection and decoding in
  {LDPC}-based space-time coded {MIMO-OFDM} systems via linear programming,''
  \emph{IEEE Trans. Signal Process.}, vol.~63, no.~13, pp. 3411--3424, Apr.
  2015.

\bibitem{wang2015diversity}
K.~Wang, W.~Wu, and Z.~Ding, ``Diversity combining in wireless relay networks
  with partial channel state information,'' in \emph{IEEE Intl. Conf. on
  Acoust., Speech and Signal Process. (ICASSP), South Brisbane, Queensland},
  Aug. 2015, pp. 3138--3142.

\bibitem{wang2016diversity}
K.~Wang and Z.~Ding, ``Diversity integration in hybrid-{ARQ} with {Chase}
  combining under partial {CSI},'' \emph{IEEE Trans. Commun.}, vol.~64, no.~6,
  pp. 2647--2659, Apr. 2016.

\bibitem{wang2018unified}
K.~Wang, ``Unified receiver design in wireless relay networks using
  mixed-integer programming techniques,'' \emph{arXiv preprint
  arXiv:1808.09649}, 2018.

\bibitem{wang2018non}
K.~Wang and Z.~Ding, ``Non-iterative joint detection-decoding receiver for
  {LDPC}-coded {MIMO} systems based on {SDR},'' \emph{arXiv preprint
  arXiv:1808.05477}, 2018.

\bibitem{wang2018iterative}
K.~Wang and Z.~Ding, ``Iterative turbo receiver for {LDPC}-coded {MIMO} systems
  based on semi-definite relaxation,'' \emph{arXiv preprint arXiv:1803.05844},
  2018.

\bibitem{wang2018integrated}
K.~Wang and Z.~Ding, ``Integrated semi-definite relaxation receiver for
  {LDPC}-coded {MIMO} systems,'' \emph{arXiv preprint arXiv:1806.04295}, 2018.

\bibitem{wang2018semidefinite}
K.~Wang, ``Semidefinite relaxation based blind equalization using constant
  modulus criterion,'' \emph{arXiv preprint arXiv:1808.07232}, 2018.

\bibitem{hochwald2003achieving}
B.~M. Hochwald and S.~Ten~Brink, ``Achieving near-capacity on a
  multiple-antenna channel,'' \emph{IEEE transactions on communications},
  vol.~51, no.~3, pp. 389--399, 2003.

\bibitem{yang2006nonlinear}
K.~Yang, J.~Feldman, and X.~Wang, ``Nonlinear programming approaches to
  decoding low-density parity-check codes,'' \emph{IEEE Journal on selected
  areas in communications}, vol.~24, no.~8, pp. 1603--1613, 2006.

\bibitem{yang2008new}
K.~Yang, X.~Wang, and J.~Feldman, ``A new linear programming approach to
  decoding linear block codes,'' \emph{IEEE Transactions on Information
  Theory}, vol.~54, no.~3, pp. 1061--1072, 2008.

\bibitem{chertkov2007pseudo}
M.~Chertkov and M.~Stepanov, ``Pseudo-codeword landscape,'' in
  \emph{Information Theory, 2007. ISIT 2007. IEEE International Symposium
  on}.\hskip 1em plus 0.5em minus 0.4em\relax IEEE, 2007, pp. 1546--1550.

\bibitem{arikan2009channel}
E.~Arikan, ``Channel polarization: A method for constructing capacity-achieving
  codes for symmetric binary-input memoryless channels,'' \emph{IEEE
  Transactions on Information Theory}, vol.~55, no.~7, pp. 3051--3073, 2009.

\bibitem{goela2010lp}
N.~Goela, S.~B. Korada, and M.~Gastpar, ``On lp decoding of polar codes,'' in
  \emph{Information Theory Workshop (ITW), 2010 IEEE}.\hskip 1em plus 0.5em
  minus 0.4em\relax IEEE, 2010, pp. 1--5.

\bibitem{arikan2008performance}
E.~Arikan, ``A performance comparison of polar codes and reed-muller codes,''
  \emph{IEEE Communications Letters}, vol.~12, no.~6, 2008.

\bibitem{feldman2004decoding}
J.~Feldman and D.~R. Karger, ``Decoding turbo-like codes via linear
  programming,'' \emph{Journal of Computer and System Sciences}, vol.~68,
  no.~4, pp. 733--752, 2004.

\bibitem{ahuja2017network}
R.~K. Ahuja, \emph{Network flows: theory, algorithms, and applications}.\hskip
  1em plus 0.5em minus 0.4em\relax Pearson Education, 2017.

\bibitem{marzetta2010noncooperative}
T.~L. Marzetta, ``Noncooperative cellular wireless with unlimited numbers of
  base station antennas,'' \emph{IEEE Transactions on Wireless Communications},
  vol.~9, no.~11, pp. 3590--3600, 2010.

\bibitem{larsson2014massive}
E.~G. Larsson, O.~Edfors, F.~Tufvesson, and T.~L. Marzetta, ``Massive {MIMO}
  for next generation wireless systems,'' \emph{IEEE communications magazine},
  vol.~52, no.~2, pp. 186--195, 2014.

\bibitem{rusek2013scaling}
F.~Rusek, D.~Persson, B.~K. Lau, E.~G. Larsson, T.~L. Marzetta, O.~Edfors, and
  F.~Tufvesson, ``Scaling up {MIMO}: Opportunities and challenges with very
  large arrays,'' \emph{IEEE signal processing magazine}, vol.~30, no.~1, pp.
  40--60, 2013.

\bibitem{lu2014overview}
L.~Lu, G.~Y. Li, A.~L. Swindlehurst, A.~Ashikhmin, and R.~Zhang, ``An overview
  of massive {MIMO}: Benefits and challenges,'' \emph{IEEE journal of selected
  topics in signal processing}, vol.~8, no.~5, pp. 742--758, 2014.

\bibitem{krishnan2014uplink}
N.~Krishnan, R.~D. Yates, and N.~B. Mandayam, ``Uplink linear receivers for
  multi-cell multiuser {MIMO} with pilot contamination: Large system
  analysis,'' \emph{IEEE Transactions on Wireless Communications}, vol.~13,
  no.~8, pp. 4360--4373, 2014.

\bibitem{shahbazpanahi2004minimum}
S.~Shahbazpanahi, M.~Beheshti, A.~B. Gershman, M.~Gharavi-Alkhansari, and K.~M.
  Wong, ``Minimum variance linear receivers for multiaccess {MIMO} wireless
  systems with space-time block coding,'' \emph{IEEE Transactions on Signal
  Processing}, vol.~52, no.~12, pp. 3306--3313, 2004.

\bibitem{wang2016robust}
K.~Wang and Z.~Ding, ``Robust receiver design based on {FEC} code diversity in
  pilot-contaminated multi-user massive {MIMO} systems,'' in \emph{IEEE Intl.
  Conf. on Acoust., Speech and Signal Process. (ICASSP), Shanghai, China}, May
  2016.

\bibitem{wang2016fec}
K.~Wang and Z.~Ding, ``{FEC} code anchored robust design of massive {MIMO}
  receivers,'' \emph{IEEE Trans. Wireless Commun.}, vol.~15, no.~12, pp.
  8223--8235, Sept. 2016.

\bibitem{wu2014cooperative}
W.~Wu, K.~Wang, Z.~Ding, and C.~Xiao, ``Cooperative multi-cell {MIMO} downlink
  precoding for finite-alphabet inputs,'' in \emph{Proc. IEEE Int. Conf.
  Acoust., Speech, Signal Process. (ICASSP), Florence, Italy}, July 2014, pp.
  464--468.

\bibitem{wu2015cooperative}
W.~Wu, K.~Wang, W.~Zeng, Z.~Ding, and C.~Xiao, ``Cooperative multi-cell {MIMO}
  downlink precoding with finite-alphabet inputs,'' \emph{IEEE Trans. Commun.},
  vol.~63, no.~3, pp. 766--779, 2015.

\bibitem{wang2017phase}
K.~Wang, L.~Jalloul, and A.~Gomaa, ``Phase noise compensation with limited
  reference symbols,'' \emph{arXiv preprint arXiv:1711.10064}, 2017.

\end{thebibliography}

\end{document}